\documentclass[preprint]{ptephy}

\preprintnumber{KYUSHU-HET-199}

\usepackage{amssymb}
\usepackage{amsmath}
\usepackage{bm}
\usepackage{mathtools}
\usepackage{comment}
\newcommand{\be}{\begin{equation}}
\newcommand{\ee}{\end{equation}}
\newcommand{\lt}{\left}
\newcommand{\rt}{\right}
\newcommand{\del}{\partial}
\newcommand{\fn}{\footnote}
\newcommand{\non}{\nonumber \\}
\numberwithin{equation}{section}
\begin{document}

\title{Renormalon structure in compactified spacetime}

\author{%
\name{\fname{Kosuke} \surname{Ishikawa}}{1},
\name{\fname{Okuto} \surname{Morikawa}}{1},
\name{\fname{Kazuya} \surname{Shibata}}{1},
\name{\fname{Hiroshi} \surname{Suzuki}}{1}, and
\name{\fname{Hiromasa} \surname{Takaura}}{1,\ast}
}

\address{%
\affil{1}{Department of Physics, Kyushu University
744 Motooka, Nishi-ku, Fukuoka, 819-0395, Japan}
\email{takaura@phys.kyushu-u.ac.jp}
}

\date{\today}

\begin{abstract}
We point out that the location of renormalon singularities in theory on a 
circle-compactified spacetime $\mathbb{R}^{d-1} \times S^1$
(with a small radius $R \Lambda \ll 1$) can differ from that on the non-compactified spacetime $\mathbb{R}^d$.
We argue this under the following assumptions, which are often realized in large $N$ theories with twisted boundary conditions:
(i) a loop integrand of a renormalon diagram is volume independent, 
i.e. it  is not modified by the compactification, 
and (ii) the loop momentum variable along the $S^1$ direction is not associated with the twisted boundary conditions
and takes the values $n/R$ with integer $n$. 
We find that the Borel singularity is generally shifted by $-1/2$ in the Borel $u$-plane,
where the renormalon ambiguity of $\mathcal{O}(\Lambda^k)$ is changed to $\mathcal{O}(\Lambda^{k-1}/R)$
due to the circle compactification $\mathbb{R}^d \to \mathbb{R}^{d-1} \times S^1$.
The result is general for any dimension $d$ and is independent of details of the quantities under consideration.
As an example, we study the $\mathbb{C} P^{N-1}$ model on $\mathbb{R} \times S^1$ 
with $\mathbb{Z}_N$ twisted boundary conditions in the large~$N$ limit.
%Our observation is crucial in considering validity of the semi-classical interpretation of the infrared renormalon.
 \end{abstract}

\subjectindex{B06, B32, B35}
%<https://publication.jps.jp/cgi-bin/ptep/submission/subject_index.cgi>
\maketitle

\section{Introduction}
% Divergence in PT and how the ambiguity cured
In perturbation theory of quantum field theory, perturbative series are typically divergent 
due to the factorial growth of the perturbative coefficients. There are two sources of this growth. 
One is the rapid growth of the number of Feynman diagrams as $\sim n!$ at the $n$th order.
The other typically originates from a single Feynman diagram whose amplitude grows factorially $\sim \beta_0^n n!$, 
and is related to the beta function of the theory, where $\beta_0$ is the one-loop coefficient of the beta function. 
The latter is known as the renormalon~\cite{tHooft:1977xjm, Beneke:1998ui}.
Such divergence of perturbative series implies that the accuracy of perturbative predictions is limited.
Through the so-called Borel procedure (which is used to obtain a finite result from divergent series),
the first source induces an imaginary ambiguity of $\mathcal{O}(e^{-2 S_I})$ 
in terms of the one-instanton action $S_I=(4 \pi)^{d/2} N/(2 \lambda)$ in $d$-dimensional spacetime, 
and the second is $\mathcal{O}(e^{-2 S_I/(N \beta_0)})$, called the renormalon ambiguity,
where $\lambda$ denotes the 't~Hooft coupling (defined as $\lambda=g^2 N$ from a conventional coupling $g$).

It is believed that the perturbative ambiguities disappear after the nonperturbative contributions are added.
It has been pointed out that the first kind of the ambiguity is canceled against the ambiguity associated with the instanton-anti-instanton calculation \cite{Brezin:1976wa,Lipatov:1976ny,Bogomolny:1980ur,ZinnJustin:1981dx}.
Here, the semiclassical configuration plays an important role, where 
the two-instanton action $2 S_I$ gives the same size of contribution 
as the first kind of perturbative ambiguity to the path integral.
On the other hand, it is not really known how the renormalon ambiguity is cured.
A clear exposition is only known in the $O(N)$ non-linear sigma model on two-dimensional spacetime,
where a nonperturbative condensate, which appears in the context of the operator product expansion, 
cancels the renormalon ambiguity \cite{David:1982qv, Novikov:1984ac, David:1983gz,Novikov:1984rf,Beneke:1998eq,Beneke:1998ui}.

%Recent attempt to cancel the renormalon ambiguity by the semiclassical object called bion
% S^1 compactification
There are currently active attempts to seek the semiclassical object which cancels the renormalon ambiguity,
expecting a scenario analogous to the cancellation mechanism of the first kind of perturbative ambiguity.
Since the instanton action is not compatible with the renormalon ambiguity by the factor $N \beta_0$, another configuration is needed.
A recent idea is to find such a configuration by the $S^1$ compactification of the spacetime 
as $\mathbb{R}^{d} \to \mathbb{R}^{d-1} \times S^1$ (see Ref.~\cite{Dunne:2015eaa} and references therein).
In order for the semiclassical calculation to be valid, the $S^1$ radius $R$
is taken to be small $R \Lambda \ll 1$, where $\Lambda$ denotes the dynamical scale.
In some theories on the compactified spacetime, a semiclassical solution that
may be able to cancel the renormalon ambiguity is found, called the bion.  
In this scenario the ambiguity associated with the bion calculation is expected to cancel the renormalon ambiguity in the theory on $\mathbb{R}^{d-1} \times S^1$
first, and then smooth connection to the theory on~$\mathbb{R}^d$ is assumed.
In Refs.~\cite{Argyres:2012vv, Argyres:2012ka, Dunne:2012ae, Dunne:2012zk, Fujimori:2018kqp},
it was claimed that the bion ambiguity is consistent with the renormalon ambiguity.   

%Does really the bion cancel the renormalon? Of great importance to examine the renormalon ambiguity clearly
So far, however, there has been no explicit confirmation that the bion truly cancels the renormalon ambiguity.
To examine the validity of the bion scenario, it is of great importance to clarify the renormalon structure on the compactified spacetime
because, as mentioned, this scenario expects the cancellation of the ambiguities due to 
the renormalon and bion first on the compactified spacetime.
The purpose of this paper is to present some insight on the renormalon structure of the theory on the compactified spacetime. 

%The understanding of the renormalon structure on the compactified spacetime is as follows.
In Ref.~\cite{Ishikawa:2019tnw}, the renormalon ambiguity in the supersymmetric $\mathbb{C}P^{N-1}$ model
on the circle-compactified spacetime $\mathbb{R} \times S^1$ with $\mathbb{Z}_N$ twisted boundary conditions
was studied.\fn{
Reference \cite{Anber:2014sda} is a pioneering work studying the renormalon in the compactified spacetime.
The authors of Ref.~\cite{Anber:2014sda} analyzed the renormalon in $SU(N)$ QCD on $\mathbb{R}^3 \times S^1$ with adjoint fermions
for small $N$, in which the bion analysis has been carried out \cite{Argyres:2012ka, Argyres:2012vv}.
On the other hand, Ref.~\cite{Ashie:2019cmy} investigated the renormalon of the same system for large $N$. 
(The analysis of Ref.~\cite{Ashie:2019cmy} utilizes the so-called large-$\beta_0$ approximation, which is a conventional tool to
analyze renormalon in non-Abelian gauge theory, combined with large $N$ limit.)}
In a systematic expansion in $1/N$,\footnote{The large $N$ limit with $R \Lambda \ll 1$ fixed was considered.}
it was found that the renormalon ambiguity of the photon condensate (and its gradient flow extension) 
changes from that in the non-compactified spacetime~$\mathbb{R}^2$. In particular, 
the Borel singularity is shifted by $-1/2$ in the Borel $u$-plane (whose definition is explained shortly) due to the compactification.

%Furthermore, a more recent work on $SU(N)$ QCD on $\mathbb{R}^3 \times S^1$ with adjoint fermions \cite{Ashie:2019cmy}, 
%in which the bion analysis has been carried out \cite{Argyres:2012ka, Argyres:2012vv},
%showed a parallel result that the renormalon ambiguity is changed due to the compactification, where
%the so-called large-$\beta_0$ approximation\fn{Roughly speaking, the large-$\beta_0$ approximation considers the so-called renormalon diagrams
%while encoding the asymptotic freedom of theory by hand.} combined with large $N$ limit is used
%to study the gluon condensate.\fn{
%The pioneering work \cite{Anber:2014sda}, which studied the same theory in the similar setup 
%to Ref.~\cite{Ashie:2019cmy},
%claimed the absence of renormalon for small $N$.}
%Both of the works of~Refs.~\cite{Ishikawa:2019tnw,Ashie:2019cmy} indicated that the Borel singularities are shifted by $-1/2$ in the Borel $u$-plane (whose definition is explained shortly) due to compactification.
%Rephrasing this, the renormalon ambiguity in~$\mathbb{R}^d$ of~$\mathcal{O}(\Lambda^k)$ is changed to
%$\mathcal{O}(\Lambda^{k-1}/R)$ in $\mathbb{R}^{d-1} \times S^1$;
%we have the renormalon ambiguity peculiar to the compactified spacetime.

\begin{figure}
\begin{center}
\includegraphics[width=9cm]{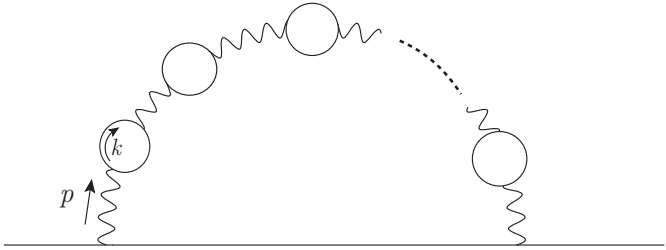}
\end{center}
\caption{Renormalon diagram.
The typical situation we consider is that the field corresponding to the wavy line satisfies the periodic boundary condition 
(and has the Kaluza--Klein (KK) momentum $p_d=n/R$), 
while the field making bubbles (solid line) satisfies the twisted boundary conditions 
(effectively corresponding to $k_d=n/(NR)$) in large $N$ theories.
The twisted boundary conditions for the field making bubbles  
are responsible for the volume independence of the loop integrand $f(p)$.}
\label{fig:1}
\end{figure}

% As a generalization of our previous work, with large-N volume independence
In this paper, as a generalization of the previous work \cite{Ishikawa:2019tnw}, we present a general mechanism 
to explain the shift of the Borel singularity, or renormalon ambiguity.
Our argument proceeds under the assumption that a loop integrand of a renormalon diagram
is not modified by the circle compactification, although we consider a sufficiently small radius $\Lambda R \ll 1$.
In other words, we assume that a loop integrand exhibits the so-called volume independence. 
This feature would be general in the large $N$ limit 
with certain twisted boundary conditions \cite{Eguchi:1982nm, Gross:1982at, Kovtun:2007py, 
Unsal:2008ch, Poppitz:2009fm, Unsal:2010qh, GonzalezArroyo:2010ss, Sulejmanpasic:2016llc}.
We also assume that the loop momentum variable of the renormalon diagram along the $S^1$ direction is given by
$n/R$ with integer $n$, and is not associated with the twisted boundary conditions.
These assumptions typically correspond to the situation explained in Fig.~\ref{fig:1}.
Under these assumptions, we study a renormalon diagram in the compactified spacetime 
$\mathbb{R}^{d-1} \times S^1$, not restricting the dimension $d$.
It tells us that the Borel singularity is shifted by $-1/2$ in the Borel $u$-plane due to the compactification,
independently of the dimension of spacetime or the details of the physical quantities under consideration.
The origin of this shift can be easily and clearly understood by effective reduction of the dimension of the momentum integration, as shall be explained.
We also treat, as an explicit example, the $\mathbb{C}P^{N-1}$ model on $\mathbb{R} \times S^1$ 
with the $\mathbb{Z}_N$ twisted boundary conditions in the large $N$ limit,
where we observe the shift of a Borel singularity for an observable defined from the gradient flow \cite{Luscher:2010iy, Luscher:2011bx}.

Let us clarify the definitions adopted in this paper to study factorially divergent series. 
For a perturbative series,
\be
\lambda \sum_{n=0}^{\infty} d_n \lt[\frac{\beta_0 \lambda}{(4 \pi)^{d/2}} \rt]^n \, ,
\ee
we define its Borel transform as
\be
B(u)=\sum_{n=0}^{\infty} \frac{d_n}{n!} u^n \, , \label{Boreldef}
\ee
and correspondingly the Borel integral is given by
\be
\frac{(4 \pi)^{d/2}}{\beta_0} \int_0^{\infty} du \, B(u) e^{-(4 \pi)^{d/2} u/(\beta_0 \lambda)} \, .
\ee
In our definition, a pole singularity of the Borel transform at $u=u_0>0$ gives 
an ambiguity in the Borel integral of order $e^{-(4 \pi)^{d/2} u_0/(\beta_0 \lambda)}=e^{-2 S_I u_0/(N \beta_0)}$.
Thus, our definition is convenient to grasp the ambiguity in terms of $e^{-S_I}$; 
it is enough to focus on the value~$u_0$
and not necessary to pay attention to the dimension $d$.
(This definition coincides with those in~Refs.~\cite{Ishikawa:2019tnw, Ashie:2019cmy}.)
As we mentioned, we show the shift of the Borel singularity by $-1/2$ in the compactified spacetime
compared to the non-compactified case.

This paper is organized as follows.
In Sect.~\ref{sec:2} we explain the general mechanism of
how the shift of the Borel singularity occurs with the circle compactification of spacetime.
%We assume volume independence of the loop integrand of a renormalon diagram.
In Sect.~\ref{sec:3}, as an example we study the $\mathbb{C} P^{N-1}$ model on $\mathbb{R} \times S^1$
with the $\mathbb{Z}_N$ twisted boundary conditions in the large~$N$ limit.
%The effective action for auxiliary fields exhibit volume independence,
%and it renders the loop integrands of the renormalon diagrams volume independent; 
%hence the above mechanism indeed works. 
The effective action of auxiliary fields exhibit volume independence, and thus, 
our large $N$ calculation essentially reduces to the one in Ref.~\cite{DAdda:1978vbw}.
We note that the above volume independence of this model has already been clarified in Ref.~\cite{Sulejmanpasic:2016llc},
although the clarification of the renormalon structure is the novel point in this paper.
Section~\ref{sec:4} is devoted to the conclusions.

\section{Renormalon structure in compactified spacetime}
\label{sec:2}

In asymptotically free theory on the non-compactified spacetime $\mathbb{R}^d$, a typical form from which a renormalon ambiguity appears 
is given by
\be
\int \frac{d^d p}{(2 \pi)^d} F(p) \lambda (p^2 e^{-C}) \, , \label{typicalinf}
\ee
where $\int d^d p$ is typically a loop integral and $C$ is a constant.
We encounter Eq.~\eqref{typicalinf} in analyzing the renormalon
using the leading logarithmic approximation, the large-$\beta_0$ approximation~\cite{Broadhurst:1993ru,Ball:1995ni,Beneke:1994qe},
and the large $N$ approximation \cite{David:1982qv, Novikov:1984ac, David:1983gz,Novikov:1984rf,Beneke:1998eq,Beneke:1998ui}. 
Here, $\lambda$ denotes the running coupling
which satisfies the renormalization group equation
\be
\mu^2 \frac{d}{d \mu^2} \lambda(\mu^2)= -\frac{\beta_0}{(4 \pi)^{d/2}}  \lambda^2(\mu^2)  \qquad {\text{with $\beta_0>0$}} \, , \label{running}
\ee
whose solution is given by
\be
\lambda(\mu^2)=\frac{(4 \pi)^{d/2}}{\beta_0} \frac{1}{\log(\mu^2/\Lambda^2)} \, ,
\ee
with a renormalization group invariant (dynamical) mass scale $\Lambda^2=\mu^2 e^{-(4 \pi)^{d/2}/[\beta_0 \lambda(\mu^2)]}$.
When the asymptotic form of $F(p)$ in the infrared (IR) region is given by\fn{
We consider a ultraviolet (UV) convergent quantity and hence the behavior of $F(p)$
in the UV region is not the same as that in the IR region.}
\be
F(p)\simeq(p^2)^{\alpha} \label{assumption2} \, ,
\ee
the Borel singularity arises at 
\be
u=\alpha+\frac{d}{2} \label{singinfvolume}
\ee 
from perturbative expansion of Eq.~\eqref{typicalinf},
which gives the renormalon ambiguity of~$\mathcal{O}(\Lambda^{2 \alpha+d})$.\fn{
This can be easily seen by repeating the subsequent argument but without compactification.}

Suppose that, in asymptotically free theory on the compactified spacetime $\mathbb{R}^{d-1} \times S^1$, we have
\be
\sum_{n=-\infty}^{\infty} \frac{1}{2 \pi R}  \int \frac{d^{d-1} \bm{p}}{(2 \pi)^{d-1}}  \, F (\bm{p}, p_d=n/R) \lambda(p^2 e^{-C}) \, , \label{start1}
\ee
as a perturbative contribution.
Here $p_d$ denotes the KK momentum along $S^1$ and is given by $p_d=n/R$, whereas 
$\bm{p}$ denotes the (continuous) momentum on $\mathbb{R}^{d-1}$.
As in Eq.~\eqref{start1}, we assume that the loop integrand/summand is not modified from the infinite-volume case~\eqref{typicalinf}.
That is, we assume volume independence of the integrand/summand.
We also assume the discrete loop momentum along the $S^1$ direction to be $p_d=n/R$ with integer $n$.\fn{
In Ref.~\cite{Ashie:2019cmy}, which studies $SU(N)$ QCD with adjoint fermions on $\mathbb{R}^3\times S^1$, 
we encounter the case that the discrete loop momentum is effectively given by $p_d=n/(NR)$ rather than $p_d=n/R$.
}
In the next section, as an example where we have Eq.~\eqref{start1}, we study the $\mathbb{C} P^{N-1}$ model on $\mathbb{R} \times S^1$ with the $\mathbb{Z}_N$ twisted boundary conditions.
It will be shown that the large $N$ limit and the $\mathbb{Z}_N$~twisted boundary conditions
play an essential role in realizing the volume independence of the loop integrand/summand~\cite{Sulejmanpasic:2016llc}.
%This feature would be general, for instance, in the large-$N$ limit with the $\mathbb{Z}_N$ twisted boundary conditions on $S^1$ direction.
%The mechanism of volume independence of the integrand/summand is explained in the next section with an explicit example
%of the $\mathbb{C} P^{N-1}$ model with the $\mathbb{Z}_N$ twisted boundary conditions.

We analyze the renormalon ambiguity involved in Eq.~\eqref{start1}.
To analyze the IR renormalon, it is sufficient to focus on the IR region by
introducing a UV cutoff $q$ to the momentum $p^2 < q^2$.
We take the UV cutoff as $\Lambda \ll q \ll R^{-1}$.
Then, due to $p^2=\bm{p}^2+(n/R)^2$, only the $n=0$ term and the range $0<\bm{p}^2<q^2$ have to be considered:
\be
\frac{1}{2 \pi R} \int_{\bm{p}^2 < q^2} \frac{d^{d-1} \bm{p}}{(2 \pi)^{d-1}} (\bm{p}^2)^{\alpha} \lambda (\bm{p}^2 e^{-C}) \, . \label{IR}
\ee 
We study the Borel transform [defined in Eq.~\eqref{Boreldef}] corresponding to Eq.~\eqref{IR},
which is obtained as\fn{
We obtain Eq.~\eqref{IRBorel} by changing the order of the momentum integration
and the infinite sum. This is indeed justified, for instance, in the region $|u| < [\alpha+(d-1)/2]/2$ (for real $u$). The result obtained in this region 
can be extended to the whole $u$-plane by analytic continuation.}
\be
B(u)=\frac{1}{2 \pi R} \int_{\bm{p}^2 < q^2} \frac{d^{d-1} \bm{p}}{(2\pi)^{d-1}}
(\bm{p}^2)^{\alpha} \lt(\frac{\mu^2 e^C}{\bm{p}^2} \rt)^{u} \, . \label{IRBorel}
\ee
Here, we note that the perturbative expansion of Eq.~\eqref{IR} in terms of $\lambda(\mu^2)$ is obtained through
\be
\lambda(p^2 e^{-C})=\lambda(\mu^2) \sum_{k=0}^{\infty} \log^k \lt(\frac{\mu^2 e^C}{p^2} \rt) \lt[ \frac{\beta_0 \lambda(\mu^2)}{({4 \pi})^{d/2}} \rt]^k \, . \label{coupexp}
\ee
The Borel transform~\eqref{IRBorel} is easily evaluated as
\be
B(u)=(\mu^2 e^C)^{u} \frac{1}{(4 \pi)^{(d-1)/2}} \frac{1}{\Gamma((d-1)/2)} \frac{1}{2 \pi R} \frac{q^{2 \alpha+d-1-2u}}{\alpha+(d-1)/2-u} \, .
\ee
This possesses a simple pole at\fn{
We assume $2\alpha+d-1>0$ for IR finiteness of Eq.~\eqref{start1}.}
\be
u=\alpha+\frac{d-1}{2} >0 \, . \label{singcompact}
\ee
The Borel singularity is shifted by $-1/2$ compared to the infinite-volume case as shown in~Eq.~\eqref{singinfvolume}. 
As a result, the renormalon ambiguity appears as
\begin{align}
&\frac{(4 \pi)^{d/2}}{\beta_0} \int_0^{\infty \times e^{\pm i \delta}} du \, B(u) e^{-{(4 \pi)^{d/2} u}/[\beta_0 \lambda(\mu)]} \non
&\sim  (\pm i \pi) \frac{1}{\beta_0} \lt(e^C\rt)^{\alpha+\frac{d-1}{2}} \frac{\sqrt{4 \pi}}{\Gamma((d-1)/2)} \frac{1}{2 \pi R} \Lambda^{2 \alpha+d-1} \, , \label{genamb}
\end{align}
where only the renormalon ambiguity is shown.
Equations~\eqref{singcompact} and \eqref{genamb} are the main results of this argument.

As one can see from Eq.~\eqref{genamb}, the renormalon ambiguity is independent of the artificial momentum cutoff $q$.
In fact, this cutoff independence holds in a broader sense.
Let us take the UV cutoff as $(\Lambda \ll) R^{-1} < q$ instead of $\Lambda \ll q \ll R^{-1}$.\fn{
However, we restrict the UV cutoff to be $q \ll Q$, where $Q$ denotes the typical scale of an observable (which is not explicitly considered here)
such that the expansion of $F$ in the low energy region is justified. ($F$ is generally a function of $p$ and $Q$.)
For instance, for the photon (or gluon) condensate defined in the gradient flow (as we consider in Sect.~\ref{sec:3}), the typical scale is $Q^2=t^{-1}$,
where $t$ is the flow time.}
In this case, we obtain the Borel transform
\begin{align}
B(u)
&=\frac{1}{2 \pi R} \int_{\bm{p}^2 < q^2} \frac{d^{d-1} \bm{p}}{(2\pi)^{d-1}}
(\bm{p}^2)^{\alpha} \lt(\frac{\mu^2 e^C}{\bm{p}^2} \rt)^{u}  \non
&\qquad
+\sum_{|n/R|<q,~ n\neq0 } \frac{1}{2 \pi R} \int_{\bm{p}^2<q^2-(n/R)^2} \frac{d^{d-1} \bm{p}}{(2 \pi)^{d-1}} (p^2)^{\alpha} \lt( \frac{\mu^2 e^C}{p^2} \rt)^u \, .
\end{align}
The first line is the same as Eq.~\eqref{IRBorel}. For the second line, since $p^2=\bm{p}^2+(n/R)^2$ always has a non-zero positive value
larger than $1/R^2 \gg \Lambda^2$,
the integrals never become singular for any $u$.
Thus, we do not have additional singularities.

As we observed,  $p_d=n/R$ with $|n| \geq 1$ does not give any renormalon singularities.
This is because the compactification radius $1/R \gg \Lambda$ plays the role of an IR cutoff for this sector. 
Hence, only the lowest KK mode (with $n=0$) can give the renormalon singularities and should be focused, 
where we have Eq.~\eqref{IR}.
This is nothing but Eq.~\eqref{typicalinf} with the replacement $d \to d-1$ (apart from the overall factor $1/(2 \pi R)$).
This replacement is the origin of the shift. Thus, the shift is simply understood as the reduction of 
the dimension of momentum integration---cf. Eqs.~\eqref{singinfvolume} and \eqref{singcompact}.

\section{Renormalon of the $\mathbb{C} P^{N-1}$ model on $\mathbb{R} \times S^1$ with $\mathbb{Z}_N$ twisted boundary conditions}
\label{sec:3} 
 
As an example where the mechanism in Sect.~\ref{sec:2} applies,
we consider the $\mathbb{C} P^{N-1}$ model on~$\mathbb{R} \times S^1$
with the $\mathbb{Z}_N$ twisted boundary conditions. The action of this model in terms of the homogeneous coordinate $z^A$ ($A=1, \dots, N$) obeying the constraint $\bar{z}^A z^A=1$
is defined by
\be
S=\int d^2 x \, \frac{N}{\lambda_0}
\lt(\del_{\mu} \bar{z}^A \del_{\mu} z^A- j_{\mu} j_{\mu} \rt) 
+S_{\rm top}  \, ,
\ee
with the current $j_{\mu}$,
\be
j_{\mu}=\frac{1}{2 i} \lt(\bar{z}^A \del_{\mu} z^A -z^A \del_{\mu} \bar{z}^A\rt) \, .
\ee
The topological term is given by
\be
S_{\rm top}=\int d^2 x \, \frac{i \theta}{2 \pi} \epsilon_{\mu \nu} \del_{\mu} j_{\nu}  \, ,
\ee
where $\epsilon_{x y}=-\epsilon_{y x}=+1$.
Here and hereafter, summation over the repeated indices is always understood. 
It is convenient to adopt the following action with auxiliary fields to carry out the large $N$ expansion \cite{Coleman:1985rnk}:
\begin{align}
S'
&=S+\int d^2 x \, \frac{N}{\lambda_0} \lt[(A_{\mu}+j_{\mu}) (A_{\mu}+j_{\mu}) + f \lt(\bar{z}^A z^A-1\rt)\rt]-\int d^2 x \, \frac{i \theta}{2 \pi} \epsilon_{\mu \nu} \del_{\mu} (A_{\nu}+j_{\nu}) \non
&=\int d^2 x \, \frac{N}{\lambda_0} \lt[-f+\bar{z}^A (-D_{\mu} D_{\mu}+f) z^A\rt] - \int d^2 x \, \frac{i \theta}{2 \pi} \epsilon_{\mu \nu} \del_{\mu} A_{\nu} \, ,
\end{align}
where $D_{\mu}=\del_{\mu}+i A_{\mu}$.
To respect the $U(1)$ gauge symmetry of the model, $A_{\mu}$ behaves as a gauge field under the transformation $z^A \to g z^A$ with $g \in U(1)$:
\be
A_{\mu} \to A_{\mu} -\frac{1}{i} g^{-1} \del_{\mu} g \, . \label{gauge}
\ee

We impose the following $\mathbb{Z}_N$ twisted boundary conditions along the $S^1$ direction for $z^A$: 
\be
z^A(x, y+ 2\pi R)=e^{2 \pi i m_A R} z^A(x,y)  \, ,
\ee 
where $(x, y) \in \mathbb{R} \times S^1$ and
\begin{align}
& m_A=\frac{A}{NR} \qquad \text{for $A=1$, \dots, $N-1$}  \, , \\
& m_N=0 \, .
\end{align}
The auxiliary fields, $A_{\mu}$ and $f$, satisfy the periodic boundary conditions.
In what follows, we analyze the renormalon ambiguity in this model by using $1/N$ expansion.
As we shall see, 
renormalon diagrams of this model possess the same integrands as the infinite-volume case.

\subsection{Volume independence of the effective action}
As pointed out in Ref.~\cite{Sulejmanpasic:2016llc}, the effective action for the auxiliary fields $S_{\rm eff}[A_{\mu}, f]$
exhibits volume independence due to the $\mathbb{Z}_N$ twisted boundary conditions and the large $N$.  
We illustrate this point, aiming for a self-contained explanation.
After integrating out $z^A$, the effective action is obtained as
\be
S_{\rm eff}[A_{\mu}, f]=-\int d^2 x \, \frac{N}{\lambda_0} f+ \sum_A {\rm Tr} \, {\rm Ln} (-D_{\mu} D_{\mu}+f) \, ,
\ee
where the topological term should be treated separately.
We first calculate the effective potential, which is obtained as $S_{\rm eff}=V_2 \cdot V_{\rm eff}(A_{\mu 0}, f_0)$;
the fields with subscript 0 denote the constant values at the saddle point;
$V_2 $ represents the volume of two-dimensional spacetime. 
$V_{\rm eff}$ is explicitly given  by
\be
V_{\rm eff}(A_{\mu 0}, f_0)=-\frac{N}{\lambda_0} f_0
+\sum_A \int \frac{d k_x}{2 \pi} \frac{1}{2 \pi R} \sum_{k_y} \ln\lt[(k_x+A_{x0})^2+(k_y+m_A+A_{y0})^2+f_0\rt] \, . \label{Veff1}
\ee
Here, the KK momentum $k_y$ is discrete:\fn{
As noted in footnote \ref{KKmode} below, the KK momentum of $z$ fields effectively reduces to $n/(NR)$
as seen from the subsequent calculations.}
\be
k_y=\frac{n}{R} \, , \qquad n \in \mathbb{Z} \, .
\ee
By using the formula
\be
\frac{1}{2 \pi R} \sum_{n=-\infty}^{\infty} F(n/R)=\sum_{n=-\infty}^{\infty} \int \frac{d k_y}{2 \pi} \, e^{i k_y 2 \pi R n} F(k_y) \, , \label{formula}
\ee
we can rewrite the infinite sum by the infinite sum of the integrals,
where the momentum shift $k_y \to k_y -m_A -A_{y0}$ is allowed.
Then, we obtain
\be
V_{\rm eff}(A_{\mu 0}, f_0)=-\frac{N}{\lambda_0} f_0
+\sum_A \sum_{n=-\infty}^{\infty} e^{-i(m_A+A_{y 0}) 2 \pi R n} \int \frac{d^2 k}{(2 \pi)^2} \, e^{i k_y 2\pi R n} \ln(k^2+f_0) \, . \label{Veff2}
\ee

It is important to note that the sum over $A$ yields
\be
\sum_A e^{- i m_A 2 \pi R n}=\sum_{j=0}^{N-1} \lt(e^{-2\pi n i/N}\rt)^j
=
\begin{cases}
N &{\text{for $n = 0$ mod $N$}} \, ,\\
0 &{\text{for $n \neq 0$ mod $N$}} \, .
\end{cases}
\ee
Thus, in the sum $\sum_{n=-\infty}^{\infty}$ in Eq.~\eqref{Veff2}, only $n$ such that $n=Nm$ with integer $m$ can contribute. 
Then, we have
\be
V_{\rm eff}(A_{\mu 0}, f_0)=-\frac{N}{\lambda_0} f_0
+N \sum_{m=-\infty}^{\infty} e^{-i A_{y 0} 2 \pi R N m} \int \frac{d^2 k}{(2 \pi)^2} e^{i k_y 2\pi R N m} \ln(k^2+f_0) \, , \label{Veff3}
\ee
where the $m=0$ term is the same contribution as the infinite-volume case, whereas the $m \neq 0$ terms are peculiar to the compactified spacetime. However, for $m \neq 0$ since we have the oscillating factor $e^{i p_y  2\pi R N m}$ in the integrand, 
these integrals vanish in the large $N$ limit where $R N \to \infty$.\fn{By applying the formula \eqref{formula} to Eq.~\eqref{Veff3},
one can see that the discrete momentum effectively reduces to $k_y=n/(NR)$ \cite{Sulejmanpasic:2016llc}.
This is the situation explained in Fig.~\ref{fig:1}.\label{KKmode}
} 
Hence, we obtain the same effective potential as the infinite-volume case~\cite{DAdda:1978vbw},
\be
V_{\rm eff}(A_{\mu0}, f_0)=V_{{\rm eff}, \infty}(A_{\mu0}, f_0)=-\frac{N}{4 \pi} f_0 \lt[\log\lt(f_0/\Lambda^2\rt)-1\rt] \, . \label{Veff4}
\ee
In Appendix \ref{app:A} we present the explicit result of the $m \neq 0$ terms and 
one can give an explicit proof that this contribution is negligible 
for large $N$ in a parallel manner to Appendix B of~Ref.~\cite{Ishikawa:2019tnw}. 
(This contribution is exponentially suppressed as $\sim e^{-N}$.)

In obtaining Eq.~\eqref{Veff4}, we apply dimensional regularization to the $m=0$ term in~Eq.~\eqref{Veff3}, where the dimension is set to be $2 \to d=2-2 \epsilon$,
and accomplish the renormalization of the bare coupling in the $\overline{\rm MS}$ scheme as
\be
\lambda_0=\lt(\frac{e^{\gamma_E} \mu^2}{4 \pi} \rt)^{\epsilon} \lambda(\mu^2) \lt[1+\frac{\lambda(\mu^2)}{4 \pi} \frac{1}{\epsilon} \rt]^{-1} \, .
\ee
The structure of the renormalization is not modified from the infinite-volume case, and the theory is indeed asymptotically free:
\be
\mu^2 \frac{d}{d \mu^2} \lambda(\mu^2)=-\frac{\beta_0}{4 \pi} \lambda^2(\mu^2) \qquad\text{with $\beta_0=1$} \, .
\ee
The $\Lambda$ scale used in Eq.~\eqref{Veff4} is defined as $\Lambda^2=\mu^2 e^{-4 \pi/[\beta_0 \lambda(\mu^2)]}$.
From Eq.~\eqref{Veff4}, the saddle point is given by
\be
f_0=\Lambda^2  \, ,
\ee
as in the infinite-volume case.
On the other hand, $A_{y 0}$ is not determined and this moduli parameter should be integrated.\fn{
The integration range is determined as follows. 
Noting that the theory is invariant under $g \in U(1)$ satisfying the non-trivial boundary condition,
\be
g(x,y+2\pi R)=e^{2 \pi i/N} g(x,y) \, ,
\ee
the shift of $A_{y}$ induced by an element $e^{i y/(RN)} \in U(1)$,
\be
A_{y} \to A_{y} -1/(RN) \, ,
\ee
reduces to an equivalent theory. Thus, the integral over $\int_0^1 d (A_{y0} R N)$ should be considered.
As long as the quantity to be integrated over this moduli parameter is independent of $A_{y0}$,
this integral has no apparent effect.}

Based on the same reasoning as above, thanks to the $\mathbb{Z}_N$ twisted boundary condition and large $N$,
the effective action for the fluctuation of the fields, $A_{\mu}=A_{\mu0}+\delta A_{\mu}$ and $f=f_0+\delta f$,
reduces to the same form as the infinite-volume case. We show it to the quadratic order~\cite{DAdda:1978vbw}:
\begin{align}
&S_{\rm eff}[\delta A_{\mu},\delta f]|_{\rm quadratic} \non
&=\frac{N}{4 \pi} \int \frac{d p_x}{2 \pi} \frac{1}{2 \pi R} \sum_{p_y}
\lt[\frac{1}{2} (p^2 \delta_{\mu \nu}-p_{\mu} p_{\nu} ) \mathcal{L}^{\delta A}_{\infty}(p) \widetilde{\delta A}_{\mu}(p) \widetilde{\delta A}_{\nu}(-p)
 -\frac{1}{2} \mathcal{L}^{\delta f}_{\infty}(p) \widetilde{\delta f}(p) \widetilde{\delta f}(-p)  \rt] \, ,  \label{Seff}
\end{align}
where we define
\begin{align}
\delta A_{\mu}(x,y)&=\int \frac{d p_x}{2 \pi} \frac{1}{2 \pi R} \sum_{p_y} e^{i p_x x +i p_y y} \widetilde{\delta A}_{\mu}(p) \, ,&
\delta f(x,y)&=\int \frac{d p_x}{2 \pi} \frac{1}{2 \pi R} \sum_{p_y} e^{i p_x x +i p_y y} \widetilde{\delta f}(p) \, ,
\end{align}
and
\begin{align}
\mathcal{L}_{\infty}^{\delta A}(p)&=\frac{2 \sqrt{p^2+4 \Lambda^2}}{p^2 \sqrt{p^2}} \log \lt(\frac{\sqrt{p^2+4 \Lambda^2}+\sqrt{p^2}}{\sqrt{p^2+4\Lambda^2}-\sqrt{p^2}} \rt)-\frac{4}{p^2} \, ,
\\
\mathcal{L}_{\infty}^{\delta f}(p)&=\frac{2}{\sqrt{p^2 (p^2+4\Lambda^2)}} \log \lt(\frac{\sqrt{p^2+4\Lambda^2}+\sqrt{p^2}}{\sqrt{p^2+4\Lambda^2}-\sqrt{p^2}} \rt) \, .
\end{align}
In Appendix~\ref{app:A}, we present the effective action including the finite-volume contributions,
which are omitted here. 
We again note that the finite-volume corrections are shown to be exponentially suppressed $\sim e^{-N}$ in the large $N$ limit
in a parallel manner to Appendix B of~Ref.~\cite{Ishikawa:2019tnw}.

\subsection{Renormalon}

To calculate the propagators of the auxiliary fields,
we add the gauge-fixing term, 
\be
S_{\rm gf}=\frac{N}{4\pi} \int \frac{d p_x}{2 \pi}  \frac{1}{2 \pi R} \sum_{p_y}  \frac{1}{2} p_{\mu} p_{\nu} \mathcal{L}_{\infty}^{\delta A}(p)
\widetilde{\delta A}_{\mu}(p) \widetilde{\delta A}_{\nu}(-p) \, ,
\ee
to the effective action Eq.~\eqref{Seff}.
Then, the propagators read
\begin{align}
\lt\langle \widetilde{\delta A}_{\mu}(p) \widetilde{\delta A}_{\nu}(q) \rt\rangle
&=\frac{4 \pi}{N} \delta_{\mu \nu} \frac{1}{p^2 \mathcal{L}_{\infty}^{\delta A}(p)} 2\pi \delta (p_x+q_x) 2 \pi R \delta_{p_y+q_y,0} \, ,
\\
\lt\langle \widetilde{\delta f}(p) \widetilde{\delta f}(q) \rt\rangle
&=-\frac{4 \pi}{N} \frac{1}{\mathcal{L}_{\infty}^{\delta f}(p)} 2\pi \delta (p_x+q_x) 2 \pi R \delta_{p_y+q_y,0} \, .
\end{align}
These are the leading-order results of the two-point functions in $1/N$.
Since they are obtained from the volume-independent effective action, 
these results are of course volume independent. It is worth noting, however, that they do not contain renormalons.\fn{
Regarding the gauge field propagator, since it is gauge dependent, this result itself does not have physical meaning.}
To see this, we consider the expansion of $1/\mathcal{L}_{\infty}^{\delta A}(p)$ and $1/\mathcal{L}_{\infty}^{\delta f}(p)$
in the high-energy region $\Lambda^2/p^2 \ll 1$ so that the perturbative expansion in the asymptotically free theory works:
\begin{align}
\frac{1}{\mathcal{L}_{\infty}^{\delta A}(p)}
&=p^2 \lt\{ \frac{\lambda(p^2 e^{-2})}{8 \pi}-\frac{\Lambda^2}{p^2} \lt[\frac{\lambda(p^2 e^{-2})}{4 \pi} +\frac{3 \lambda^2(p^2 e^{-2}) }{16 \pi^2} \rt] +\mathcal{O}(\Lambda^4/p^4)\rt\} \, ,
\\
\frac{1}{\mathcal{L}_{\infty}^{\delta f}(p)}
&=p^2 \lt\{ \frac{\lambda(p^2)}{8 \pi} +
\frac{\Lambda^2}{p^2} \lt[\frac{\lambda(p^2)}{4 \pi}-\frac{\lambda^2(p^2)}{16 \pi^2}\rt]+\mathcal{O}(\Lambda^4/p^4) \rt\} \, .
\end{align}
Since $\Lambda^2 =\mu^2 e^{-4\pi/[\beta_0 \lambda(\mu^2)]}$ is zero in perturbative evaluation,
these quantities are evaluated in perturbation theory (PT) as\fn{
The coefficients of $(\Lambda^2/p^2)^k$ are regarded as Wilson coefficients, and they are calculated in perturbation theory
as they are given by perturbative series in $\lambda$.}
\be
\lt. \frac{1}{\mathcal{L}_{\infty}^{\delta A}(p)} \rt|_{\rm PT}=p^2 \frac{\lambda(p^2 e^{-2})}{8 \pi} \,, \qquad \lt. \frac{1}{\mathcal{L}_{\infty}^{\delta f}(p)} \rt|_{\rm PT}=p^2 \frac{\lambda(p^2)}{8 \pi} \, . \label{PropPT}
\ee
These results do not contain renormalon divergence; they are truncated at $\mathcal{O}(\lambda)$.
If one uses a general renormalization scale $\mu$ in accordance with the concept of fixed-order perturbation theory,
the infinite sums in $\lambda(\mu^2)$ appear through Eq.~\eqref{coupexp} but they can be unambiguously resummed.

Renormalons appear when the propagator containing the running coupling, as in Eq.~\eqref{PropPT}, 
is involved in a loop integrand.
As simple examples, let us consider the condensates, $\langle f(x) f(x) \rangle$ and $\langle F_{\mu \nu}(x) F_{\mu \nu}(x) \rangle$:
\begin{align}
\langle f(x) f(x) \rangle &=\Lambda^4-\frac{4 \pi}{N} \int \frac{d p_x}{2 \pi} \frac{1}{2 \pi R} \sum_{p_y} \lt. \frac{1}{\mathcal{L}_{\infty}^{\delta f}(p)} \rt|_{p^2 < q^2} \, ,
\\
\langle F_{\mu \nu}(x) F_{\mu \nu}(x) \rangle
&=\frac{4 \pi}{N} \int \frac{d p_x}{2 \pi} \frac{1}{2 \pi R} \sum_{p_y} \lt. \frac{2}{\mathcal{L}_{\infty}^{\delta A}(p)} \rt|_{p^2 < q^2} \, . \label{FF}
\end{align}
Since the condensates are UV divergent, we introduce a UV cutoff $q$ to define them.
The perturbative evaluations of these quantities are given by\fn{
The perturbative results of the condensates are the same as the loop integrations of
the perturbative expressions of the integrands. We also note that the perturbative expression is reliable only in 
the high-energy region.
Using such a result in the low energy-region, which is not justified, is a cause of a renormalon ambiguity.}
\begin{align}
\langle f(x) f(x) \rangle|_{\rm PT} &=-\frac{4 \pi}{N} \int \frac{d p_x}{2 \pi} \frac{1}{2 \pi R} \sum_{p_y} \lt. p^2 \frac{\lambda(p^2)}{8 \pi} \rt|_{p^2<q^2, \text{expansion in $\lambda(\mu)$}} \, ,
\\
\langle F_{\mu \nu}(x) F_{\mu \nu}(x) \rangle|_{\rm PT}
&=\frac{4 \pi}{N} \int \frac{d p_x}{2 \pi} \frac{1}{2 \pi R} \sum_{p_y} \lt. p^2 \frac{\lambda(p^2 e^{-2})}{4 \pi}\rt|_{p^2<q^2, \text{expansion in $\lambda(\mu)$}}^{} \, ,
\label{photoncond}
\end{align}
where we explicitly show that their integrands should be expanded in $\lambda(\mu)$.\fn{
If the integrands are not expanded in $\lambda(\mu)$, the integrals are ill-defined since they contain the poles 
around $p^2 \sim \Lambda^2$, which are related to the renormalon ambiguities.}
Now, we indeed encounter Eq.~\eqref{start1}, assumed in the general argument in Sect.~\ref{sec:2};
the loop integrands are not modified from the infinite-volume case, but the integration measure is modified to that of the compactified 
spacetime. We note that since the auxiliary fields $A_{\mu}(x)$ and $f(x)$ satisfy the periodic boundary condition,
the momentum along the $S^1$ direction is given by $p_y=n/R$ and has nothing to do with the twisted boundary conditions.
Following the result in Sect.~\ref{sec:2}, the renormalon ambiguities are given by
\begin{align}
\langle f(x) f(x) \rangle|_{\text{renormalon}}&=\mp i \pi \frac{1}{N} \frac{1}{2 \pi R} \Lambda^{3} \, ,
\\
\langle F_{\mu \nu}(x) F_{\mu \nu}(x) \rangle|_{\text{renormalon}} 
&=\pm i \pi \frac{2 e^{3}}{N} \frac{1}{2 \pi R} \Lambda^{3}  \, .
\end{align}
As already noted in Sect.~\ref{sec:2}, the renormalon ambiguities are independent of the UV cutoff.
These renormalon ambiguities are peculiar to the compactified spacetime, 
since they  depend on $R$.

To give an example of a UV-finite observable which possesses a renormalon ambiguity, 
we consider the gradient flow \cite{Luscher:2010iy,Luscher:2011bx}. 
The flow equation is given by
\be
\del_t B_{\mu}(t,x)=\del_{\nu} G_{\nu \mu}(t,x)+\alpha_0  \del_{\mu} \del_{\nu} B_{\nu}(t,x) \, , \qquad B_{\mu}(t=0,x)=A_{\mu}(x) \, ,
\ee
where $G_{\mu \nu}(t,x)=\del_{\mu} B_{\nu}(t,x)-\del_{\nu} B_{\mu}(t,x)$ is the field strength of the flowed gauge field;
$\alpha_0$~is a constant regarded as a gauge parameter;
$t$ is called the flow time, whose mass dimension is $-2$.
The flowed gauge field is obtained as 
\begin{align}
&B_{\mu}(t,x)
\non
&=A_{\mu0}+\int d^2 x' \int \frac{d p_x}{2 \pi} \frac{1}{2 \pi R} \sum_{p_y} e^{i p (x-x')} \lt[\lt(\delta_{\mu \nu}-\frac{p_{\mu} p_{\nu}}{p^2}\rt)e^{-tp^2} +\frac{p_{\mu} p_{\nu}}{p^2} e^{-\alpha_0 t p^2}\rt] \delta A_{\mu}(x') \, .
\end{align}
Using the flowed gauge field, we can construct an observable $\langle G_{\mu \nu}(t,x) G_{\mu \nu}(t,x) \rangle$,
\be
\langle G_{\mu \nu}(t,x) G_{\mu \nu}(t,x) \rangle
=\frac{4 \pi}{N} \int \frac{d p_x}{2 \pi} \frac{1}{2 \pi R} \sum_{p_y}   \frac{2}{\mathcal{L}_{\infty}^{\delta A}(p)} e^{-2 t p^2} \, , \label{GG}
\ee
where the Gaussian damping factor makes this quantity UV finite.
In perturbation theory, it is given by
\be
\langle G_{\mu \nu}(t,x) G_{\mu \nu}(t,x) \rangle|_{\rm PT}
=\frac{4 \pi}{N} \int \frac{d p_x}{2 \pi} \frac{1}{2 \pi R} \sum_{p_y} \lt. p^2 \frac{\lambda(p^2 e^{-2})}{4 \pi} e^{-2 t p^2} \rt|_{\text{expansion in $\lambda(\mu)$}} \, .
\ee
To analyze the renormalon in this quantity, we introduce a UV cutoff $\Lambda^2 \ll q^2 \ll t^{-1}$ as done in Sect.~\ref{sec:2}.
Since the integrand has the same behavior as that of the photon condensate \eqref{photoncond} in the IR region due to $e^{-2 t p^2} \simeq 1$,
we have the same renormalon ambiguity as $\langle F_{\mu \nu}(x) F_{\mu \nu}(x) \rangle$:
\be
\langle G_{\mu \nu}(t,x) G_{\mu \nu}(t,x) \rangle|_{\text{renormalon}} 
=\pm i \pi \frac{2 e^{3}}{N} \frac{1}{2 \pi R} \Lambda^{3} \, . \label{renGG}
\ee 
As seen from the above reasoning, it is fairly general that 
the leading renormalon ambiguity of the photon (or gluon) condensate defined by the gradient flow, which is a UV-finite observable,
is the same as that of the photon (or gluon) condensate defined with the UV cutoff.

In Eq.~\eqref{renGG}, only the leading renormalon ambiguity of~$\langle G_{\mu \nu}(t,x) G_{\mu \nu}(t,x) \rangle$ is shown.
By considering the expansion of $e^{- 2 t p^2}$ in $t p^2$ at higher order, 
we obtain the renormalon ambiguity beyond this order of the form
\be
\langle G_{\mu \nu}(t,x) G_{\mu \nu}(t,x) \rangle|_{\text{renormalon}} 
=\pm i \pi \lt(c_0 \frac{1}{R} \Lambda^3+c_1 t \frac{1}{R} \Lambda^5+c_2 t^2 \frac{1}{R} \Lambda^7+\dots \rt) \, ,
\ee
according to the argument in Sect.~\ref{sec:2}, where $c_0$, $c_1$, $c_2$, \dots denote the constants. 
In the Borel $u$-plane, these renormalon ambiguities correspond to
the singularities at $u=3/2$, $5/2$, $7/2$, \dots. 
These positions are different from the infinite-volume case, where the singularities are located at
$u=2$, $3$, $4$, \dots.

We finally note that the emergence of the renormalon ambiguities is indeed 
an artifact of perturbation theory.
By seeing Eq.~\eqref{GG}, which is not evaluated in perturbation theory,
one can see that this quantity is unambiguous because
any divergence is not found in this expression.
It indicates that the renormalon ambiguities found above are cured after
nonperturbative effects are properly added.

\section{Conclusions}
\label{sec:4}

% Summary, effective action のvolume independenceはrenormalonのvolume independenceを意味しない
In this paper, we have presented a general argument that the renormalon structure is significantly affected by the circle compactification
of the spacetime as $\mathbb{R}^d \to \mathbb{R}^{d-1} \times S^1$ with small $S^1$ radius $R \Lambda \ll 1$.
The assumptions of this argument are that
(i) a loop integrand of a renormalon diagram\fn{
Here, we mean by a renormalon diagram that its loop integrand possesses a one-loop running coupling.}
is not modified by the compactification, and that 
(ii) the discrete loop momentum along the $S^1$ direction of the renormalon diagram is not associated with 
the twisted boundary condition of the system and is given by $n/R$ with integer $n$.
%The property (i) is often realized in the large $N$ theories with twisted boundary conditions,
%while the property (ii) implies that the loop momentum is not associated with the twisted boundary conditions.
Under these assumptions, we showed that a shift of the renormalon singularity generally occurs 
due to the circle compactification $\mathbb{R}^d \to \mathbb{R}^{d-1} \times S^1$. 
In particular, the singularity is shifted by $-1/2$ in the Borel $u$-plane regardless of the dimension of spacetime $d$
and details of the quantities under consideration.
This can be easily understood by the reduction of the dimension of 
the loop momentum integral, which stems from the fact that only the lowest KK mode 
can give renormalon singularities. 

% Our previous work(SUSY CP^N )はこれの例 
As an example, we studied the $\mathbb{C} P^{N-1}$ model on $\mathbb{R} \times S^1$ with the $\mathbb{Z}_N$
twisted boundary conditions in the large $N$ limit.
In this model, the above properties (i) and (ii) are indeed realized.
The shift by $-1/2$ in the Borel $u$-plane was explicitly shown 
by studying the photon condensate which is defined by the gradient flow and is a UV-finite observable.
As already mentioned,
the previous work, Ref.~\cite{Ishikawa:2019tnw}, which studied the supersymmetric $\mathbb{C} P^{N-1}$ model on $\mathbb{R} \times S^1$,
had provided examples where this mechanism applies in the large $N$ approximation.
%and $SU(N)$ QCD with adjoint fermions on~$\mathbb{R}^3 \times S^1$.\fn{
%The latter model requires a slight modification.
%This is because this model partially breaks the gauge symmetry due to compactification 
%and the effective action is not completely same as that on~$\mathbb{R}^4$.
%However, the difference can be removed by a shift of loop momentum of a renormalon diagram.}

Finally, we emphasize that the volume independence of the effective action does {\it not} always indicate the volume independence
of the renormalon structure, as we observed in the example of the $\mathbb{C} P^{N-1}$ model.
In this model, the volume-independent effective action gave the two-point functions or propagators
which do  not contain renormalon ambiguity.
The renormalon ambiguity arises when these propagators (determined from the effective action)
are included as loop integrands. 
Such quantities do not show volume independence any more, and the renormalon structure is {\it not} kept intact.

\begin{comment}
We finally argue the fate of the renormalon ambiguity in the $\mathbb{C} P^{N-1}$ model.
As seen from Eq.~\eqref{FF}, the condensate itself can be evaluated as a finite quantity
when we use the leading result in $1/N$.
Thus, it is clear that the renormalon ambiguity is artifact of perturbation theory.
Hence, the renormalon ambiguity found in the first term should be canceled by the nonperturbative term
represented by the second term in the following equation:
\begin{align}
\langle F_{\mu \nu}(x) F_{\mu \nu}(x) \rangle
&=\frac{4 \pi}{N} \int \frac{d p_x}{2 \pi} \frac{1}{2 \pi R} \sum_{p_y}  
 \frac{2}{\mathcal{L}_{\infty}^{\delta A}(p)} \bigg|_{{\text{PT}},~p^2 < q^2} \non
 &\qquad +\frac{4 \pi}{N} \int \frac{d p_x}{2 \pi} \frac{1}{2 \pi R} \sum_{p_y}  
 \lt( \frac{2}{\mathcal{L}_{\infty}^{\delta A}(p)} -\frac{2}{\mathcal{L}_{\infty}^{\delta A}(p)} \bigg|_{\text{PT}} \rt) \bigg|_{p^2 < q^2} \, .
\end{align}
Hence, the usual OPE scenario extensively investigated in 1980's indeed works
with the compactification of the spacetime.
\end{comment}

%
 
\section*{Acknowledgments}
The authors are grateful to Akira Nakayama for fruitful discussions.
They also thank Toshiaki Fujimori, Tatsuhiro Misumi, and Norisuke Sakai for discussions.
This work was supported by JSPS Grants-in-Aid for Scientific Research
numbers JP18J20935 (O.M.), JP16H03982 (H.S.), and~JP19K14711 (H.T.).

\appendix

\section{Finite volume corrections}
\label{app:A}
We present the effective potential and effective action which contain the finite volume corrections.

The effective potential is given by
\be
V_{\rm eff}(A_{\mu0}, f_0)=V_{{\rm eff}, \infty}(A_{\mu0},f_0)+V_{{\rm eff}, \rm{finite}}(A_{\mu0},f_0) \, ,
\ee
with the finite volume correction
\be
V_{\rm eff, \rm{finite}}(A_{\mu0}, f_0)=-\frac{N}{\pi} \sum_{m \neq 0} e^{- i A_{y0} 2 \pi R N m} \frac{\sqrt{f_0}}{2 \pi R N |m|} K_1(\sqrt{f_0} 2 \pi R N |m|) \, ,
\ee
where $K_{\nu}(z)$ denotes the modified Bessel function of the second kind.

The effective action including the finite volume effects is given by
\begin{align}
S_{{\rm eff}}
&=\frac{N}{4 \pi} \int \frac{d p_x}{2 \pi} \frac{1}{2 \pi R} \sum_{p_y} 
\non
&\qquad\times
\biggl\{\frac{1}{2} (p^2 \delta_{\mu \nu}-p_{\mu} p_{\nu}) \lt[\mathcal{L}_{\infty}^{\delta A}(p)+\mathcal{L}_{\rm finite}^{\delta A}(p) \rt] \widetilde{\delta A}_{\mu}(p) \widetilde{\delta A}_{\nu}(-p)  \non
&\qquad\qquad
-\frac{1}{2} \lt[\mathcal{L}_{\infty}^{\delta f}(p)+\mathcal{L}_{\rm finite}^{\delta f}(p) \rt] \widetilde{\delta f}(p) \widetilde{\delta f}(-p) \non
&\qquad\qquad\qquad
-\lt(\delta_{\mu y} -\frac{p_{\mu} p_{y}}{p^2}\rt) \mathcal{L}_{\rm finite}^{\rm mix}(p)\lt[\widetilde{\delta A}_{\mu}(p) \widetilde{\delta f}(-p)+\widetilde{\delta f}(p) \widetilde{\delta A}_{\mu}(-p)\rt] \biggr\}
\end{align}
with
\begin{align}
\mathcal{L}_{\rm finite}^{\delta A}(p)
&=\int_0^1 d x \sum_{m \neq 0} e^{-i A_{y0} 2 \pi R N m} e^{i x p_y 2 \pi R N m}  (2x-1)^2\non
&\qquad\qquad\times \frac{2 \pi R N |m|}{\sqrt{\Lambda^2+x(1-x)p^2}} 
K_1 (\sqrt{\Lambda^2+x(1-x)p^2 } 2 \pi R N |m|) \non
&\qquad
-\frac{2}{i p_y} \int_0^1 d x \sum_{m \neq 0} e^{-i A_{y0} 2 \pi R N m} e^{i x p_y 2 \pi R N m} (2x-1) \non
&\qquad\qquad\qquad\times 2 \pi R N m K_0(\sqrt{\Lambda^2+x(1-x)p^2 } 2 \pi R N |m|) \, ,
\\
\mathcal{L}_{\rm finite}^{\delta f}(p)
&=\int_0^1 d x \sum_{m \neq 0} e^{-i A_{y0} 2 \pi R N m} e^{i x p_y 2 \pi R N m} \non
&\qquad\qquad\times \frac{2 \pi R N |m|}{\sqrt{\Lambda^2+x(1-x)p^2}} 
K_1 (\sqrt{\Lambda^2+x(1-x)p^2 } 2 \pi R N |m|) \, ,
\end{align}
and
\begin{align}
&\mathcal{L}_{\rm finite}^{\rm mix}(p)=i \int_0^1 d x \sum_{m \neq 0} e^{-i A_{y0} 2 \pi R N m} e^{i x p_y 2 \pi R N m} 2 \pi R N m
K_0 (\sqrt{\Lambda^2+x(1-x)p^2 } 2 \pi R N |m|) \, .
\end{align}
We note that these results are consistent with the gauge invariance; 
the tensors before $\widetilde{\delta A}_{\mu}$ are transverse.

\bibliographystyle{utphys}
\bibliography{BibforCPN}

%%%%%%%%%%%%%%%%%%%%%%%%%%%%%%
\begin{comment}

\end{comment}
%%%%%%%%%%%%%%%%%%%%%%%%%%%%%%

\end{document}